\documentclass[
twocolumn,
superscriptaddress,
 amsmath,amssymb,
prb,
floatfix,
]{revtex4-1}

\usepackage{graphicx}
\usepackage{dcolumn}
\usepackage{bm}
\usepackage{epstopdf}
\usepackage{hyperref}

\begin{document}

\title{Bundling dynamics regulates the active mechanics and transport in \\
carbon nanotube networks}
\author{Myung Gwan Hahm$^{1,2}$, Hailong Wang$^3$, Hyunyoung Jung$^1$, Sanghyun Hong$^1$, Sung-Goo Lee$^4$, Sung-Ryong Kim$^5$, Moneesh Upmanyu$^3$, Yung Joon Jung$^1$\\
\vspace{0.1in}
\footnotesize $^1$Department of Mechanical and Industrial Engineering, Northeastern University, Boston, MA 02115\\
\footnotesize $^2$Department of Mechanical Engineering and Materials Science, Rice University, Houston, Texas, 77005 USA\\
\footnotesize $^3$Group for Simulation and Theory of Atomic-scale Material Phenomena (stAMP), Department of Mechanical and Industrial Engineering, Northeastern University, Boston, MA 02115\\ 
\footnotesize $^4$Information and Electronics Polymer Research Center, Korea Research Institute of Chemical Technology, Deajeon 305-600, Republic of Korea\\
\footnotesize $^5$Department of Polymer Science and Engineering, Chungju National University, Chungbuk 380-702, Republic of Korea
}


\begin{abstract}
High-density carbon nanotube networks (CNNs) continue to attract interest as active elements in nanoelectronic devices, nanoelectromechanical systems (NEMS) and multifunctional nanocomposites. The interplay between the network nanostructure and the its properties is crucial, yet current understanding remains limited to the passive response. Here, we employ a novel superstructure consisting of millimeter-long vertically aligned singe walled carbon nanotubes (SWCNTs) sandwiched between polydimethylsiloxane (PDMS) layers to quantify the effect of two classes of mechanical stimuli, film densification and stretching, on the  electronic and thermal transport across the network. The network deforms easily with increase in electrical and thermal conductivities suggestive of floppy yet highly reconfigurable network. Insight from atomistically informed coarse-grained simulations uncover an interplay between the extent of lateral assembly of the bundles, modulated by surface zipping/unzipping, and the elastic energy associated with the bent conformations of the nanotubes/bundles. During densification, the network becomes highly interconnected yet we observe a modest increase in bundling primarily due to the reduced spacing between the SWCNTs. The stretching, on the other hand, is characterized by an initial debundling regime as the strain accommodation occurs via unzipping of the branched interconnects, followed by rapid re-bundling as the strain transfers to the increasingly aligned bundles. In both cases, the increase in the electrical and thermal conductivity is primarily due to the increase in bundle size; the changes in network connectivity have a minor effect on the transport. Our results have broad implications for filamentous networks of inorganic nanoassemblies composed of interacting tubes, wires and ribbons/belts.
\end{abstract}

\maketitle

The need for scalable synthesis of robust carbon-nanotube based devices has sparked the recent interest in high density carbon nanotube networks (CNNs). A diverse range of applications have been successfully realized and include flexible nanoelectronics~\cite{cnt:Avouris:2007, cnt:KocabasRogers:2005, cnt:HurRogers:2005, ntc:JungKarAjayan:2006, cnt:CaoRogers:2008}, chemical and mechanical sensing~\cite{nt:KongDai:2000, nt:StampferHierold:2006, nt:LeeHong:2011}, heat transport~\cite{nt:HoneZettl:1999, cnt:ItkisHaddon:2007}, multifunctional membranes~\cite{nt:SrivastavaAjayan:2004, nt:MeiBaughman:2005, nt:YamadaHata:2010}, supercapacitors~\cite{nt:DuPan:2005} and nanocomposites~\cite{ntc:AjayanRubio:2000}. While it is well-known that the properties of the CNNs degrade in comparison to those of the individual single-walled nanotubes (SWCNTs), their expression is actively controlled by changes in the underlying filamentous nanostructure and its interplay with the network properties.  Current understanding remains limited to the passive response. For example, both structural rigidity of and transport across these networks vary dramatically with the degree of alignment and the density of junctions between the nanotubes~\cite{nt:HoneFischerSmalley:2000, cnt:KocabasRogers:2007, cntr:BerhanSastry:2004, nt:SomuUpmanyu:2010}. Little is known regarding the active interplay between the nanoscale structure within these assemblies and their properties, yet it is of critical importance in almost all envisioned applications. Of particular interest is the response to mechanical stimuli and its effect on the electronic transport~\cite{nt:MeiBaughman:2005, nt:YamadaHata:2010}. The mechanics is expected to be non-linear, similar in principle to the mechanics of nanotube suspensions~\cite{nt:IslamYodh:2004} as well as filamentous assemblies of semi-flexible biopolymers such as actin, collagen and fibrin~\cite{bio:MacintoshJanmey:1995, fil:Gardel:2004, bio:StormMacintoshJanmey:2005}. Indeed, recent studies on severely deformed and physically interconnected, aligned nanotube networks have uncovered a thermally stable, viscoelastic response~\cite{cnn:XuHata:2010}. Unlike these and the biopolymer networks, though, the as-grown and/or assembled CNNs that we study here are not physically cross-linked and the SWCNTs can easily slide at the inter-tube junctions. In that sense, these as-grown CNNs represent an extreme limit wherein filaments of unprecedented rigidity interact via considerably weaker van Der Waals forces. 
In this article, we focus on the mechanics of aligned CNNs, assembled into a novel superstructure, which allows us to directly quantify the interplay between mechanics and transport in these networks. We complement the experiments with coarse-grained computations that uncover the underlying nanoscale mechanisms.
\begin{figure*}[thb]
\centering
\includegraphics[width=1.5\columnwidth]{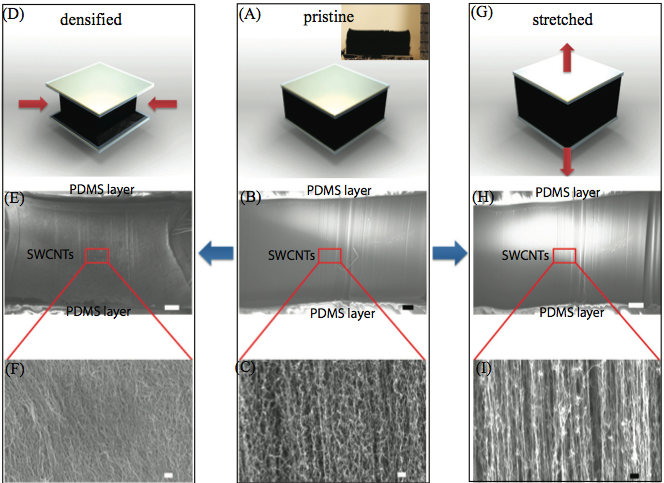}
\caption{Nanostructural characterization of the CNNs with the S/CNN/S superstructures. (A, D, G) Schematic illustration of the (A) as-fabricated (labelled ``pristineÓ, here and elsewhere), (D) densified and (G) stretched network. Ti/Au metal pads are integrated into the PDMS layers which contact the nanotube terminations, as indicated. The imposed densification and stretching strains are approximately the same, $\epsilon\approx10\%$. (inset, A) Side view of the forest of vertically-aligned SWCNTs that are used to assemble the superstructure. (B-C) Low- and high-resolution SEM images of the aligned CNN within the pristine S/CNN/S superstructures. (B-C) Cross-sectional SEM images, (B) low-resolution and and (C) high-resolution, of the assembled superstructure showing the as-grown vertically aligned SWCNTs capped by PDMS layers. (E-F, H-I) Same as in (B) and (C), but for densified and stretched superstructures. The scale bars are $200\,\mu$m and $100$\,nm for the low- and high-resolution SEM images, respectively.}
\label{fig:fig1}
\end{figure*}

\section{As-synthesized and strained SWCNT networks}
The architectural motif for our study is a sandwich structure consisting of millimeter-long vertically aligned SWCNTs sandwiched between two PDMS thin films, henceforth referred to as S/CNN/S. The superstructure is shown schematically in Fig. 1A; the inset shows the side view of the vertically-aligned forest of SWCNTs synthesized using an ethanol based chemical vapor deposition (CVD) technique. The entire assembly is capped by PDMS layers through a multi-step contact polymer transfer method. The interplay between mechanics and electronic transport requires characterization before and after deformation, and to this end, we have integrated nanometer thick metal contact pads onto the PDMS layers that are in direct contact with the nanotube terminations;  see Methods and Supplementary Information (SI) for additional details on the fabrication process.

The as-grown SWCNTs are firmly embedded into the PDMS layers (SI, Fig. S3), a natural consequence of the strong PDMS- SWCNT interfaces~\cite{ntc:AjayanRubio:2000}. The relatively long lengths of the SWCNTs is crucial as it ensures that the effects of the PDMS layers are minimized at equilibrium and also during their active response. Additionally, each constituent nanotube in the vertically aligned network makes contact with both PDMS layers. The chief benefit of the fully percolated network is that the PDMS-SWCNT contact area remains constant during deformation. While the absence of any interrupted nanotubes is a departure from the nanostructures that we expect in typical CNNs~\cite{fil:UpmanyuBarber:2005}, their effect on the overall network connectivity is minimized since i) the density of these interruptions scales inversely with the SWCNT length and is exceedingly small for the millimeter-sized nanotubes employed here, and ii) the effect on the mechanics is limited to strains beyond which the individual nanotubes become extended; these are quite large since the as-grown vertically aligned SWCNTs are intrinsically wavy.

Figures~\ref{fig:fig1}B and~\ref{fig:fig1}C show low- and high-resolution SEM images of the assembled CNNs within the superstructure. Macroscopically, the SWCNTs are aligned, yet this is not strictly true at the nanoscale, evident from wavy nanostructure seen in the SEM images (Fig.~\ref{fig:fig1}C). The filaments are largely bundles of nanotubes that are periodically rippled and also interconnected with an average inter-bundle spacing of $r\approx10$\,nm. The variations in the nanotube curvature and their lateral assembly that contributes to the quenched disorder is quite typical of the dense nanostructures associated with as-grown vertically aligned carbon nanotube forests and has its origins in the strongly confined and spatio-temporally non-uniform catalytic growth of the individual nanotubes~\cite{cntr:Terrones:1997, cntr:ZhuAjayan:2002, nt:NessimHartThompson:2009}.

We investigate the active response of the assembled S/CNN/Ss with respect to two classes of mechanical stimuli, densification and stretching, as shown schematically in Figs.~\ref{fig:fig1}D and~\ref{fig:fig1}G. For the former, the entire S/CNN/S is immersed in isopropyl alcohol (IPA) and then dried via evaporation (Methods). The eventual densification occurs in two stages: i) an initial surface tension driven elastocapillary aggregation during immersion as well as evaporation, and ii) further densification, mainly during evaporation, in which the bundles on average stick together due to the van der Waals forces once the lateral distance the between the bundles approaches the inter-graphitic spacing~\cite{cnt:ZhengchunAjayan:2009, cnt:ZhouzhouHartLu:2010, cnt:TawfickHart:2011}. The net strain due to densification is $\epsilon\approx10\%$ and the aligned CNN is transformed into rigid structure shown in Fig.~\ref{fig:fig1}E. The high-resolution image (Fig.~\ref{fig:fig1}F) shows that while the CNN is still textured, it is heavily interconnected with a slightly larger average bundle size.

Our approach to impose stretching strains on the CNNs is fairly straightforward and illustrated in Fig.~S2. The capping layers are pulled apart in a strain-controlled fashion using simple straining apparatus. The strain is applied to SiO$_2$ wafers that are retained on the capping PDMS layers post-transfer for this specific reason. The CNNs are deformed to a maximum displacement of $250\,\mu$m, or a uniaxial strain that is approximately the same as that during densification, $\epsilon\approx10\%$.\footnote{Note that the strain represents the imposed strain on the SiO$_2$ layers.  The load and the resultant strain transferred to the CNNs is expected to be different as some of the strain accommodation occurs within the PDMS layers and the PDMS/SiO$_2$ interfaces.}
The CNN stretches easily, characteristic of a floppy network that deforms primarily via soft modes. The lateral compression is negligible, i.e. the Poisson's ratio $\nu\approx0$. Unlike the viscoelastic response of physically interconnected CNNs~\cite{cnn:XuHata:2010}, there is no strain recovery following deformation. The forces required to deform the network are quite small and we therefore do not extract the force-displacement curve and limit the characterization, both structural and the transport, to the maximum imposed strain, $\epsilon\approx10\%$. The low magnification image shows little change, yet the nanostructure is quite different (Fig.~\ref{fig:fig1}I). The network topology becomes increasingly aligned and there is an increase in bundle size and the network porosity. The frequency of the interconnections between the bundles is also reduced.

\section{Transport Characteristics}
\subsection{Electro-mechanically coupled response} 
\begin{figure}[htb]
\centering
\includegraphics[width=\columnwidth]{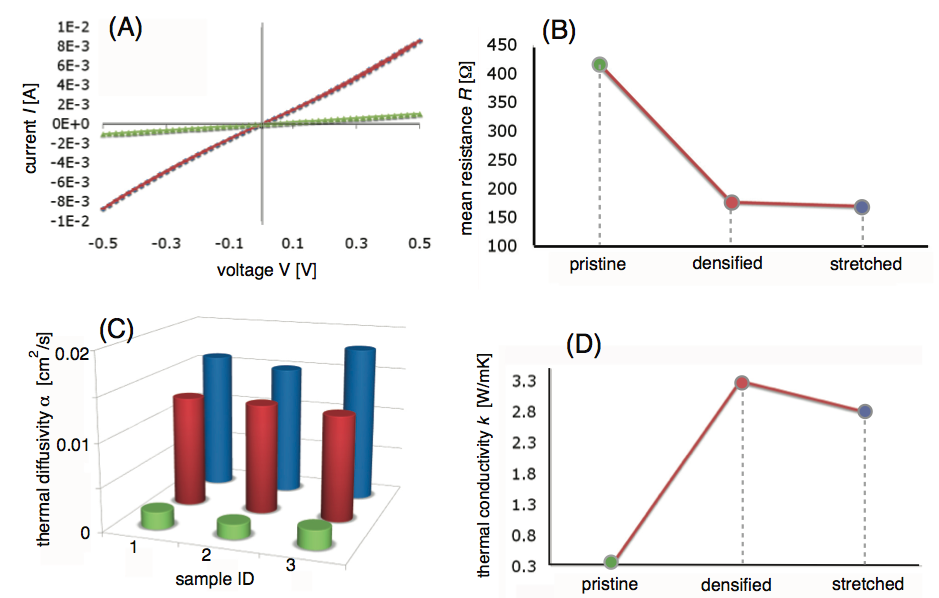}
\caption{Electrical and thermal characterization of the pristine (green), densified (red) and stretched (blue)  S/CNN/S superstructures. (A) The measured current-voltage ($I-V$) curves. The curves for strained superstructures are almost identical. (B) The mean resistance is averaged over seven fabricated superstructures (see SI). The errors bars are smaller than the solid circles and not shown. (C) Histograms for the thermal diffusivities $\alpha$ for the pristine and modified samples, for three different fabricated S/CNN/S superstructures. (D) Average thermal conductivities of pristine, densified and stretched SWCNT networks.}
\label{fig:fig2}
\end{figure}

Two-terminal current- voltage ($I-V$) characteristics of the pristine and deformed superstructures are extracted in order to quantify the electro-mechanically coupled response. The curves for one of the assembled superstructures is shown in Fig.~\ref{fig:fig2}A. The response is linear for the entire voltage regime ($\pm0.5$\,V) and representative for each of the fabricated samples. The reciprocal of the slope is the total resistance, sum of the contributions from the CNNs - their intrinsic resistance, $R_{int}$ - and that of the contacts, $R_c$. The mean resistance, $\bar{R}$, averaged over seven different fabricated samples, is plotted in Fig.~\ref{fig:fig2}B. Since the number of nanotubes that contact the metal pads is conserved during the two deformation modes, the effect of the mechanical stimuli on the contact resistance can be safely ignored. Then, the variation we see is due to the change in the intrinsic resistance of the CNNs. It is roughly the same for both modes and is substantially reduced compared to the pristine CNNs. The strain factor, a quantitative measure of the coupled response, is ($\Delta\bar{R}/\bar{R})/\epsilon\approx6$. Unlike the resistance change, the intrinsic strain factor of the CNNs will be higher as the extracted value is modified by the contact resistance. 

\subsection{Thermo-mechanically coupled response} The room temperature thermal characterization is based on thermal diffusivities extracted for pristine and modified CNNs. A steady-state temperature profile is established across the metal pads and directly yields the diffusivity $\alpha$ (see Methods). Figure~\ref{fig:fig2}C shows the extracted thermal diffusivities for three different fabricated S/CNN/S superstructures. In each case, the diffusivity decreases; a modest reduction for densified CNNs (34\% on average) while an order of magnitude reduction for the stretched CNNs. The thermal conductivity, the product of the diffusivity and the volumetric heat capacity $k = (\rho C_p)\alpha$ where $\rho$ and $C_p$ are the network density and specific heat respectively, is a more direct measure of the thermo-mechanically coupled response and is plotted in Fig.~\ref{fig:fig2}D for the three different CNNs. The conductivity of the pristine CNNs is well below those of individual nanotubes and their crystalline ropes~\cite{nt:HoneZettl:1999, cnt:BerberKwonTomanek:2000, nt:HoneFischerSmalley:2000} even after we account for the volume filling fraction of the network. The significant reduction is characteristic of the strong phonon-phonon Umklapp scattering that is highly sensitive to the nanostructure, in particular its network connectivity~\cite{nt:HoneZettl:1999}. More importantly, the order of magnitude increase in the mean thermal conductivity suggests that the imposed deformation suppresses the  scattering, either due to changes in network connectivity and/or the extent of bundling. 
The thermal factor associated with the thermo-mechanical coupling, ($\Delta\bar{k}/\bar{k})/\epsilon\approx1$. 

\section{Coarse-grained Computations}
The coupling between CNN mechanics and transport indicates an interplay between network density, the quenched disorder and extent of lateral assembly of the individual nanotubes, which we now fully explore using experimental-scale computations. All-atom calculations are clearly prohibitive due to the involved time- and length-scales. 
The inter-tube interactions are evidently central to accurately capturing the overall network response. To this end, we develop a coarse-grained approach wherein i) the individual nanotubes are approximated as linear chain of beads connected by springs that can bend, twist and stretch and ii) the van der Waal inter-tube interactions are extracted self-consistently using the universal graphitic potential integrated over the surface conformations of the individual nanotubes~\cite{cntr:Girifalco:2000, cntr:LiangUpmanyu:2005b, cntr:Buehler:2006}. The stiffness of the springs that connect the beads are tailored to the elastic properties of the individual nanotubes, extracted from atomic-scale simulations and available as radius dependent empirical expressions~\cite{cntr:ZhigileiSrivastav:2005}. For simplicity, we assume that the entire network is composed of nanotubes of the same chirality and therefore radius, (8,8) SWCNTs with radius $r=0.543$\,nm. The inter-tube interaction is the sum of the interaction between beads on different nanotubes is modeled as a 6-12 Lennard-Jones pair potential. 
Classical many-body dynamics technique is employed to generate the conformations of the SWCNTs within the initial equilibrated structure and to follow their evolution during the two classes of mechanical stimuli~\cite{book:AllenTildesley:1989}. The resultant spring-bead approximation allows access to the structural and energetic evolution of a relatively large network consisting of millions of atoms and over several microseconds.

\begin{figure}[htb]
\centering
\includegraphics[width=\columnwidth]{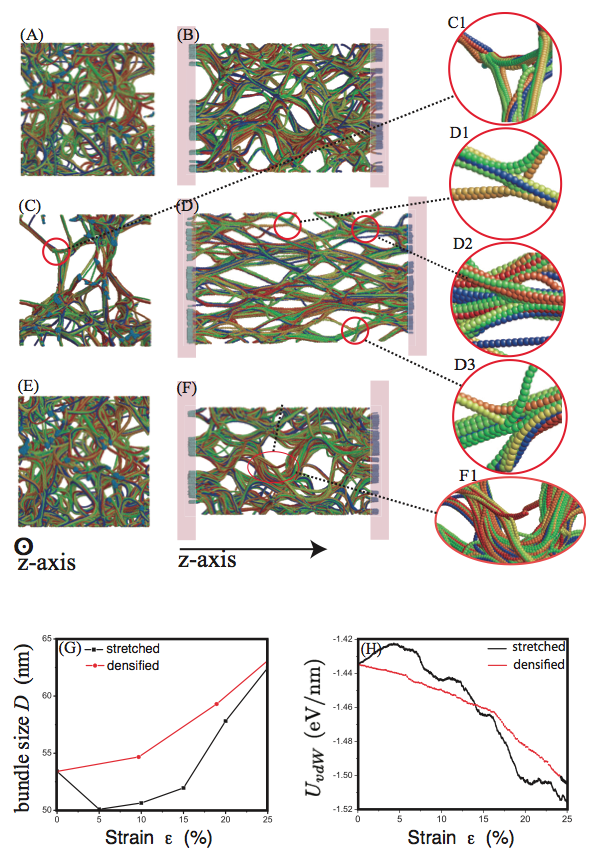}
\caption{The strain dependence of the network nanostructure in the computations. (A-E) Top (left) and side (right) views and of the coarse-grained network nanostructure extracted from simulations performed at $273^\circ$K, for (A-B) the pristine, relaxed network, (C-D) network subject to $\epsilon= 25\%$ uniaxial stretching strain, and (E-F) the densified network. The coloring scheme is based on the nanotube identity of each discretized element, i.e. bead. The length of each SWCNT is $400$\,nm, the dimensions of the systems are $200$\,nm$\times$$200$\,nm$\times$$400$\,nm, and the average distance between SWCNTs is $10$\,nm. The circled regions highlight representative mechanisms associated with the strain accommodation and their magnified view is also shown in panels C1, D1-D3 and F1. (G-H) Topology and energy change induced by deformation. (G) Evolution of the maximum bundle size $D_{max}$ and (H) the inter-tube energy $U_{vdW}$ as function of strain during stretching and densification.}
\label{fig:fig3}
\end{figure}

Figures~\ref{fig:fig3}A and~\ref{fig:fig3}B shows the top and side views of the coarse-grained (bead) structure of the relaxed, percolated network that is part of a much larger CNN. The precursor as-constructed CNN is composed of 400 SWCNTs, each $\approx400$\,nm long and spaced $10$\,nm apart to match the inter-bundle spacing observed in the experiments. We fix the terminal ends of the nanotubes, a simple approximation of the strong PDMS-SWCNT interfaces that stabilize the individual SWCNT ends within the S/CNN/S superstructure. The quenched disorder is generated by subjecting each nanotube to an initial compressive strain of $\approx25\%$, a measure of the extent of disorder, and then dynamically relaxing the entire network to a local energy minimum.\footnote{This is clearly a simplified approximation of the disorder in vertically-aligned SWCNTs which are likely curved intrinsically due to defects and related inhomogeneities during growth. However, as we show later, the approximations in the computations are sufficient to capture the salient mechanisms that control the mechanical response of the CNNs.} The SWCNTs immediately assemble into randomly branched and interconnected bundles composed of wavy and entangled conformations, quite like the network structure seen in the experiments. The lateral assembly is significant; the maximum bundle size of the network after relaxation is $D_{max}=53$\,nm. Further assembly is limited by the elastic energy cost associated with bending, twisting and stretching the individual tubes; the computations allow us to directly quantify the energetic interplay between gain in the van Der Waal energy due to their self-assembly $U_{vdW}$, and the elastic energy cost, $U_e$. 

Top and side-views of the coarse-grained nanostructure within the stretched and densified networks ($\epsilon=25\%$) are shown in Figs.~\ref{fig:fig3}C-\ref{fig:fig3}D and \ref{fig:fig3}E-\ref{fig:fig3}F, respectively. The strain rate during the deformation is held constant, $\approx10^2/\mu$s. The network is not allowed to compress laterally as in the experiments ($\nu\approx0$). The network porosity evolves as expected - decreases during stretching and increases during densification - yet the extent of the assembly visibly increases. We quantify the overall changes in the network by monitoring the two main parameters associated with its nanostructure, i) the maximum bundle size $D_{max}$ over multiple cross-sectional ($XY$) slices through the network (see Methods), and ii) the stored inter-tube interaction energy, $U_{vdW}$ , which is directly available. Their strain dependence is plotted in Fig.~\ref{fig:fig3}G-\ref{fig:fig3}H. During densification, the bundle size increases monotonically and non-linearly and provides quantitative confirmation of the enhanced bundling observed in the experimental images (Fig.~\ref{fig:fig1}F). The increase in the magnitude of the inter-tube energy (Fig.~\ref{fig:fig3}H) is again consistent with lateral assembly of the bundles. We observe a similar increase in the bundle size and $|U_{vdW} |$ in the stretched CNNs but the temporal trends show a slight departure during early stages of deformation ($\epsilon\le 5\%$). Specifically, the network initially debundles; the average bundle size reduces by $\approx7\%$ at an applied strain of $\epsilon\le 5\%$ and $|U_{vdW}|$ accordingly decreases.

The network topology following the two deformation modes is qualitatively different, as seen in Figs.~\ref{fig:fig3}C-\ref{fig:fig3}F. During stretching, the entangled conformations of the individual nanotubes are ironed out during the initial debundling regime. The strains are accommodated by the interconnections that connect the branch points between the entangled bundles. As a result, the bundles themselves are effectively shielded from the applied load and the interconnections are on average pulled apart that causes the bundles to dissemble. Past a critical point, the load transfers to the bundles and they start to assemble again, also evident from the trends in the inter-tube energy (Fig.~\ref{fig:fig3}H). More importantly, the bundles become increasingly aligned, which facilitates their assembly, and the network connectivity decreases. 

The insets in (Figs.~\ref{fig:fig3}C1 and~\ref{fig:fig3}D1-\ref{fig:fig3}D3) highlight some of the nanoscale mechanisms that are involved in the network reconfiguration during stretching. Figure~\ref{fig:fig3}C1 and~\ref{fig:fig3}D2 show the zipping together of two smaller bundles into a larger bundle. The exact opposite, i.e. unzipping, can be seen in Fig.~\ref{fig:fig3}D1 wherein an interconnecting double strand (lower left) unzips from the surface of the bundle as it accommodates the imposed stretch. The combination of the two mechanisms, surface unzipping of a SWCNTs (colored green) and a double strand (colored yellow) and the zipping together of the two bundles, can be seen in Fig.~\ref{fig:fig3}D3. We do not observe substantial slip of the nanotubes and/or bundles since the network reconfiguration occurs mainly via soft modes.

During densification, the increasingly bent conformations of the SWCNTs and the bundles must also accommodate the decrease in the average inter-bundle distance. At sufficiently large strains, the combination of lateral confinement and the branched network topology drives the lateral assembly. There is a concomitant increase in the alignment and network connectivity relative to the pristine network, visible from the network configuration shown in Figs.~\ref{fig:fig3}E-\ref{fig:fig3}F. The network reconfigures via surface zipping and unzipping at the bundles, similar to the nanoscale mechanisms in effect during stretching. The enhanced network density forces the nanotubes to pack more efficiently without dramatic increase in the elastic cost, and a dominant mechanism by which this is realized is the twisting of the bundles; one such representative configuration is shown in Fig.~\ref{fig:fig3}F1. In essence, the twist allows the network to store the additional increase in the energy stored by increasing the effective inter-tube interaction area~\cite{cntr:LiangUpmanyu:2005b}.

\begin{figure}
\centering
\includegraphics[width=0.9\columnwidth]{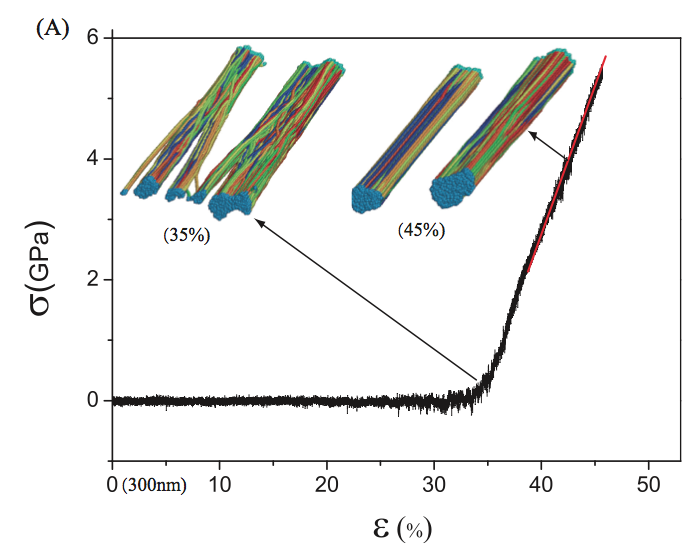}
\caption{\label{fig:fig4} Engineering stress-strain curve for the CNNs following stretching. The inset shows the evolution of the total energy of the system. The snapshots of the network at large strains are also shown ($\epsilon=35\%$ and $\epsilon=45\%$).}
\end{figure}

Figure~\ref{fig:fig4} shows the stress-strain curve during stretching of the CNNs. The initial regime indicates a soft network with a very small yet positive YoungÕs modulus, $E\approx20\pm10$\,MPa. This is characteristic of a floppy network in that the connectivity is below the rigidity threshold for this class of systems. The strain accommodation is primarily through the changes in network connectivity, lateral assembly and alignment, clearly visible in the network configuration corresponding to $\epsilon=35\%$ (left inset). The modulus and the stress can also become negative locally, an indication of transients associated with a self-healing network. At larger strains $\epsilon>35\%$, the distance between the fixed ends approaches the length of the SWCNTs and the network is no longer floppy. The Young's modulus is of the order of $\sim50$\,GPa (indicated by the linear fit in Fig.~\ref{fig:fig4}); the network configuration at $\epsilon=45\%$ (right inset) reveals an almost fully formed bundle. 

\section{Discussion and Conclusions}
Our results challenge the traditional view of CNNs as a static network of carbon nanotubes and their bundles. Rather, a detailed analysis of the active mechanical response of the CNNs reveals a dynamic network that reconfigures its nanostructure and the network topology via the extent of lateral assembly of the SWCNTs. This additional degree of freedom modifies the network mechanics as it changes the distribution of the bundle diameters and therefore their stiffness. Clearly, existing network models are ill-equipped to handle these strongly anisotropic and non-affine deformations modes and the relevant theoretical modifications are addressed elsewhere [Wang and Upmanyu, in preparation]. The trends in the network reconfiguration during stretching and densification shed light on the electro- and thermo-mechanical couplings, which we now discuss in detail. The enhanced bundling of the CNNs is the key to the observed increase in conductivities, in particular its effect on the electronic transport. Firstly, it is well-known that the electron dispersion perpendicular to the bundle axis due to the inter-tube coupling decreases the separation of the conduction and valence bands around the Fermi level, in particular the non-degenerate bands. The exact nature of the shift depends on the chirality of the nanotubes; it is as high as 0.25\,eV for the bundles of (6,6) and (8,4) SWCNTs~\cite{cntr:KwonSaitoTomanek:1998,cntr:StahlLengeler:2000}. Secondly, the doubly degenerate states also split within the bundle. As a classical example, strong band-splitting causes the (10, 0) nanotubes to become metallic upon bundling~\cite{cnt:ReichOrdejon:2002}. The combination of these two effects reduces the band gap in semiconducting SWCNTs. Of course, the reverse is also true for metallic nanotubes, especially for armchair nanotubes, but the typically higher ratio of semiconducting SWCNTs (the intrinsic electrical heterogeneity~\cite{cnt:Hersam:2008}) in the as-grown forests leads to enhanced metallicity upon bundling. Lastly, the phonon transport that determines the thermal conductivity also increases with bundling due to a non-linear increase in the number of transport channels along the bundle axis~\cite{nt:HoneZettl:1999}. Interestingly, the non-monotonic variation in the extent of bundling observed during stretching of the CNNs (Fig.~\ref{fig:fig3}G) implies that the network initially undergoes a reduction in its conductivity during the debundling regime. Since this occurs at relatively small strains ($\epsilon\le5\%$), it was not observed during the electrical and thermal characterization of the S/CNN/S superstructures.

The individual SWCNTs are also strained during overall deformation of the CNN, further modifying their band-gap. The strains, a combination of stretching, bending and twisting, can arise directly upon load transfer to the individual nanotubes, and indirectly due to the conformations of the individual SWCNTs at the branch points and within the bundles. For example, a net twist within a bundle occurs at the expense of bending and twisting of the individual tubes~\cite{cntr:LiangUpmanyu:2005b}. Again, both chirality and form of the strain modify the sign and extent of change in the band-gap, yet due to the higher density of semiconducting nanotubes in the heterogeneous forests, we expect the overall band-gap to become smaller. The conductivities are also modified by the network topology, in particular the extent of branching. In non-percolated networks, this is a major contribution as the metallic nanotubes are effectively shielded with increasing alignment~\cite{nt:SomuUpmanyu:2010}. The effect is diminished here due to percolation, yet the branch points in a network serve as scattering sites for electronic and phonon transport. Once stretched well past the initial debundling regime, the topology evolves such that the branch-point density reduces and further contributes to the increase in the thermal and electrical conductivity across the CNN. During densification, though, the network becomes increasingly branched as the nanotubes bend to accommodate the strains. Since the extent of bundling also increases, the overall increase in the conductivities is then primarily due to bundling and the debilitating effects due to scattering at the branch points are relatively less important.

In summary, the combination the experimental characterization and experimental-scale coarse-grains computations clearly show that the mechanics of  carbon nanotube networks is actively regulated by lateral assembly. Both densification and stretching result in enhanced bundling, although the network connectivity evolves differently. The increase in both electrical and thermal conductivities indicate that the bundling has a dominant effect on the overall electron and phonon transport through the network. The extracted strain and thermal factors for the CNNs that we focus here are not exceptionally large yet have technological relevance because of the large strains that can be easily absorbed by the network. Although we have not explored the effect of average length of the SWCNTs, it is intuitive that increased lengths will enhance the connectivity as the individual SWNCTs can bend and twist more easily and the bundling will become more important. Our results also identify network scale features that can be engineered for controlled transport through these networks and have broad implications for a wide range of inorganic and organic filamentous networks composed of assemblies of nanotubes, nanowires, and nanobelts. 

\section{Materials}
\subsection{Synthesis of vertically aligned SWCNTs} A multilayered substrate (Al/SiO$_2$/Si) is employed for the synthesis of the millimeter long SWCNTs. A 20\,nm thick Al buffer layer is first deposited on a SiO$_2$($100$\,nm)/Si wafer. The surface is coated with a $0.7$\,nm thick Co catalyst layer using an e-beam evaporator and then placed inside a quartz tube. The tube is evacuated to $\approx15$\,mTorr and then exposed to a $5\%$~H$_2$-95\%~Ar mixture carrier gas at a pressure of 700~Torr and temperature $850^\circ$C. After the stabilization, high-purity anhydrous ethanol ($99.95\%$) vapor is supplied from the bubbler as a carbon source which results in the growth of vertically-aligned SWCNTs.

\subsection{Assembly of the sandwiched hybrid superstructure, S/CNN/S} A two step polymer casting method is use to synthesize the S/CNN/S superstructure. Each step is illustrated schematically in SI (Fig.~S1) 
A bilayer metal contact pad (Ti($5$\,nm )/Au($150$\,nm)) is first deposited on the top-side of the as-grown vertically aligned SWCNTs using a sputter coater (Fig.~S1a). A spin coater is used to coat the SiO$_2$ wafer with liquid phase PDMS (weight ratio of $10:1$) and then $\approx80\%$ cured. The top-side of the CNN with the metal pad is placed on the surface of the PDMS coating and then fully cured at $110^\circ$C for 90 seconds (Fig.~S1b). The SiO$_2$ wafers are detached from bottom  (Fig.~S1c and~S1d) and the bilayer metal pad is deposited on the bottom-side of the CNN (Fig.~S1e). It is finally capped with the second PDMS layer by pouring liquid PDMS and then curing at $110^\circ$C for two hours (Fig.~S1f-h). 

\subsection{Electrical and Thermal Chracterization} A  Janis ST-500 electric probe station connected to a Keithley 240 sourcemeter was employed for electrical characterization. The thermal conductivities of the three different SWCNTs networks were measured by using standard laser flash thermal constant analyzer (UlVAC TC7000- HNC).

\subsection{Coarse-grained computations} The radius dependent stiffness for SWCNTs are based on prior atomic simulation~\cite{cntr:ZhigileiSrivastav:2005}. All stiffness used in our computations are for $(8, 8)$ SWCNTs (stretching $k_s$ = 626.46\,eV/{\AA}, bending $k_b$ = 9063.02\,eV/{\AA} and torsional $k_t$ = 6252\,eV/{\AA}). The spring length connecting two neighbor beads has uniform length $\Delta l = 2$\,nm and is small enough such that the orientation dependence of the interaction between beads can be ignored. The cohesive energy ($-0.845$\,eV/nm) and the equilibrium inter-tube distance (1.4\,nm) between parallel (8, 8) SWCNTs are readily available from the universal graphitic potential and serve as the two parameters for fitting the coarse-grained Lennard-Jones pair potential, $\sigma=1.534$\,nm and $\epsilon=0.81$\,eV~\cite{cntr:LiangUpmanyu:2005b, cntr:Buehler:2006}. The CNN is generated as described in the text and is stretched in a strain-controlled fashion by uniformly displacing the fixed ends of the SWCNTs (Z-direction). The Poisson's ratio in the experiments is negligible to simulate this, the cross-section area ($X-Y$ plane) of the periodic computational cell is held constant. Densification is simulated by laterally compressing the computational cell while maintaining the distance between the fixed ends of the SWCNTs. The time step for the dynamics is fixed, $\Delta t=5$\,fs. To capture the changes in the network topology during physical loading, we track the bundle the maximum bundle size by sectioning the network into 1000 cross sections. Within each section, two SWCNTs are deemed to belong to one bundle if the inter-tube distance is smaller than the cutoff  distance between the average first and second neighbors in an idealized hexagonally packed SWCNT bundle.

\end{document}